# Spectroscopic Characterization of Metallocene Single Crystals Grown by Physical Vapor Transport Method

*Ian B. Logue[1], Sandaruka Jayasooriya Arachchilage[1], Lance M. Griswold[1], Moses B. Gaither-Ganim[1], Lincoln W. Weber[1], Robyn Cook[2], Stephen Hofer[1], Praveena Satkunam, Dipanjan Mazumdar[1], Poopalasingam Sivakumar[1\*], and Bumsu Lee[1\*]*

[1]Southern Illinois University, 1263 Lincoln Dr., Carbondale, IL 62901, USA

[2]Cottey College, 1000 W Austin Blvd, Nevada, MO 64772

[*]Correspondence: bumsu.lee@siu.edu, psivakumar@siu.edu

**Abstract**: High-quality metallocene single crystals with a low density of impurities and high homogeneity were prepared using the physical vapor transport method. These crystals were then characterized using various spectroscopic tools and X-ray diffraction. Laser-induced breakdown spectroscopy confirmed the presence of metal ions in each freshly grown sample despite all these crystals undergoing physical deformation with different lifetimes. X-ray diffraction analysis confirmed that all our metallocene single crystals retained a monoclinic structure at room temperature. The vibrational properties of our metallocene crystals were examined using Raman and Fourier-transform infrared spectroscopy. The inter- and intra-ring vibrational modes, along with additional modes associated with the crystalline form, were identified as inherent vibrational properties of our metallocene single crystals. Given the increasingly important role of metallocene in organic solar cells, organic light-emitting displays and molecular quantum systems, this research will enhance our understanding of the intrinsic physical properties of cleaner, more crystalline metallocene single crystals.

## 1. Introduction

A metallocene molecule is an organometallic compound in which a single metal ion is sandwiched between two cyclopentadienyl rings. Since the study on metallocene started with the discovery of bis(cyclopentadienyl)iron, also known as ferrocene, nearly 70 years ago [1], they have been extensively used in various places from catalysts and fuel additives [2], pharmaceuticals [3-4], and even in liquid crystal technologies [5]. Research on metallocene compounds has focused primarily on vapor and solution phases and some crystals obtained from a solution growth method [6-10]. In recent years, these materials have gained renewed attention for their new role in organic photovoltaic cells [11], organic light emitting displays [12] and new host platforms for molecular quantum information [13]. Therefore, there is an increasing need to understand clean metallocene solids to exploit their intrinsic physical properties. Here, we used high-quality metallocene single crystals grown by the physical vapor transport (PVT) method to examine their inherent physical properties. Our self-assembled, free-standing metallocene crystals are highly homogeneous and have a low impurity density. This is achieved by separating light impurities from the crystallization zone through a secondary purification process during crystal growth [14]. This method will provide an optimal matrix for measuring the intrinsic properties of these materials without interference from unwanted impurities, defects, or disorders.

This study presents the structural and vibrational characterization of three types of metallocene single crystals (ferrocene, nickelocene and cobaltocene), prepared using the PVT method. Several characterization tools were employed to study on our single crystal samples, including X-ray diffraction (XRD), laser-induced breakdown spectroscopy (LIBS), Raman spectroscopy, and Fourier transform infrared (FTIR) spectroscopy.

## 2. Results and Discussion

Figures 1a-c show photos of ferrocene, nickelocene, and cobaltocene crystals. These crystals typically grow into bar-shaped structures with nice facets. However, depending on the PVT growth conditions, thin and wide crystals may also be present in the batch. The ferrocene, nickelocene, and cobaltocene appear orange, dark green, and dark purple, respectively, to the naked eye. The molecular structure of metallocene is displayed in figure 1d.

To identify the metal ions in the metallocene molecular crystals, we applied LIBS, a powerful analytical technique for determining the elemental composition of materials. This allowed us to confirm the metal ions present within the crystals and determine whether molecular decomposition occurred during crystal growth process. This method involves directing a short, high-intensity laser pulse at a sample surface, which causes ablation and the formation of a high-temperature plasma containing excited atoms, ions, and free electrons. As the plasma cools, the excited species relax to lower energy levels, emitting light at characteristic wavelengths corresponding to the elements present in the sample [15,16]. LIBS spectra obtained from our ferrocene, nickelocene, and cobaltocene crystals are compared in figure 2. Prominent atomic emission lines at wavelengths of 234.35 nm, 239.56 nm, 258.59 nm, 261.19 nm and 274.75 nm were observed in the LIBS spectrum of our ferrocene crystals, confirming the presence of iron (Fe) inside. Similarly, emission lines characteristic of nickel (Ni), including 218.55 nm, 220.67 nm, 231.60 nm, 239.5 nm and 251.60 nm, were detected in our nickelocene crystals. Cobalt (Co) emission lines at 219.36 nm, 230.79 nm, 241.16 nm, 253.34 nm and 258 nm also confirm the presence of cobalt in our cobaltocene single crystals. All of these LIBS spectra successfully identified the presence of Fe, Ni, and Co metal ions in each metallocene single crystal, demonstrating that metallocene molecules are maintained in crystalline form without undergoing molecular deformation during crystal growth.

XRD measurements were next performed to characterize the structural properties of our metallocene crystals. Metallocene crystals are generally known to prefer monoclinic crystal structure at ambient conditions, although triclinic or orthorhombic forms have also been observed under certain temperature and pressure conditions [17,18]. At room temperature, the crystal structures of ferrocene and nickelocene have been found to belong to the $P2_1/n$ monoclinic cell symmetry with the $C_{2h}^5$ factor group, where two centrosymmetric metallocene molecules are positioned at the corners and center of the unit cell [8,19]. The crystal structure of our metallocene crystals were presented in the inset of figure 3a. The well-defined parallelogram surfaces are known to be found in our metallocene crystals, as shown in figure 1, confirming the facet defined by Miller indices of (110), (1$\bar{1}$0) and (001) of the monoclinic crystal structure [20,21]. We selected single crystals with thin, wide facets to provide enough surface area for the XRD beams to focus on the uniform, highly homogeneous surface of crystals. This maintains the consistency and quality of the crystals for our XRD measurements. Since a single piece of the crystal was scanned on its wide and uniform basal plane, the exhibited XRD peaks confirm a (00n) direction of a monoclinic $P2_1/a$ (*unique axis b*) structure of our metallocene crystals, albeit with slightly different lattice constants (see the inset of each figure 3). Our XRD results showed that differences in the molecular composition of our metallocene crystals do not result in conformational differences within the symmetry group, and that the van der Waals packing energies of the metallocene molecules are quite similar, as predicted in a previous theoretical study [22].

The following studies examine the vibrational properties of our metallocene single crystals, employing two techniques: Raman spectroscopy and FTIR spectroscopy. Earlier research has explained that ferrocene molecules in the gas form have a prismatic structure (eclipsed, $D_{5h}$ point group), whereas those in crystalline form have rings arranged in an antiprismatic configuration (staggered, $D_{5d}$ point group) [20]. With the point symmetry of $C_i$, the factor group of a ferrocene crystal with a $D_{5d}$ group of the molecule is $C_{2h}^5$ [22]. The molecule in the $D_{5h}$ system coupled with the $C_i$ symmetry is estimated to have 34 available vibrations, while 57 fundamental vibrational modes are expected in the $D_{5d}$ system, with 15 Raman active and 10 infra-red (IR) active vibrations in both cases [20,23]. However, in the crystalline form, some degeneracies are eliminated, revealing the fundamental modes of vibration that were previously forbidden in the molecule within the symmetry group of the crystal structure [20,24,25]. For example, it is commonly known that molecular vibrations with even parity (*g*) are Raman active, whereas those with odd parity (*u*) are IR active, but, in crystals, degeneracy can be lifted, and forbidden symmetries also allow in Raman and IR transitions [20,24]. In addition, peak separation due to the Davydov splitting is also expected since two metallocene molecules in the unit cell exist [26]. Under this factor group of a ferrocene crystal, the vibrational spectrum is expected to include both intra-ring and inter-ring (metal-ligand) vibrations of individual ferrocene molecules, as well as intermolecular vibrations from the crystalline form.

Figure 4a shows the Raman spectra of freshly grown single crystals of ferrocene, nickelocene and cobaltocene, which were measured at room temperature. Fundamentally, the aromatic cyclopentadienyl rings exhibit vibrational modes that are similar to those of benzene molecules [23,27]. Consequently, vibrational modes associated with ring deformations, such as C-C stretching and C-H bending or stretching, are predominantly observed in metallocene compounds. The vibrational spectra are also expected to include ring tilt and the stretching and bending modes of the ring-metal ligand. These inter- and intra-ring modes are found within the range of 180 ~ 3300 cm$^{-1}$. In our figure 4, a ferrocene crystal exhibited the clearest vibrational modes of the three metallocene crystal types due to its negligible background fluorescence and greater stability in an ambient environment. This result supports previous findings that ferrocene crystals are the most stable metallocene crystals under ambient conditions. Unlike ferrocene crystals, a significant fluorescence background is present in the Raman spectra of nickelocene and cobaltocene crystals, even though the laser excitation energy ($\lambda_{exc}$= 785 nm) used in the Raman spectroscopy was much below the absorption edge of the samples. This emission may be related to the near-infrared luminescence, as demonstrated in a previous cobaltocene study [28], or to phosphorescence resulted by defect states formed during the measurement. Therefore, we primarily focused our discussion on the Raman spectra measured in our ferrocene single crystals. As previously mentioned, 15 Raman-active modes related to inter- and intra-ring vibrations are expected in the metallocene samples with $C_{2h}^{5}$ symmetry and these modes were clearly identified in our study of ferrocene single crystals. Among them, 11 Raman modes were consistently observed in our crystals, as was the case in previous studies conducted on ferrocene solutions [23]. Additional 4 modes ($\nu_2, \nu_{14}, \nu_{27}$, and $\nu_{28}$) were found only in the crystal, which are enabled by the additional crystalline symmetry, relating to the C-H bending ($\perp,\parallel$) and ring distortion ($\perp,\parallel$) [20,29,30]. Interestingly, the stretching modes of $\nu_4$ and $\nu_{16}$, which are derived from metal-ring ligands, depend highly on the type of metallocene, whereas the ring vibration modes are not. This is because the strength of the metal-ring bond varies greatly depending on the type of metal ion present in the metallocene, which contrasts with the bonds in the cyclopentadienyl ring [23]. As the separation between the metal and the ring increases in the order nickelocene > cobaltocene > ferrocene, the of $\nu_4$ and $\nu_{16}$ modes in our Raman spectra also exhibit the same order in Raman energy with a clear distinction. This contrasts with the other intra-ring modes, which demonstrate only minor changes in Raman energy. Another interesting point is that the $\nu_{28}$ vibration of the $E_{2g}$ ring distortion ($\perp$), which is known to be absent in its vapor or solution form, was evidently present in all of our crystal samples. Their Raman mode energies differ significantly, which confirms once again that the strength of the metal-ligand bond varies depending on the type of metallocene molecule.

It is worth noting that the previous study had expected the inactive $\nu_7$ mode in the molecular form to be visible in the crystalline form [20], but no $\nu_7$ peak was observed in our metallocene crystals. In addition, we followed the assignment of 1356 cm$^{-1}$ peak as $\nu_{26}$ as suggested by the previous studies [23,30]. Other vibrations related to inter-molecular (lattice) vibrations of metallocene crystals are expected to be found at frequencies below 100 cm$^{-1}$ [31], but these modes were not included in our study due to the limited measurement range of our Raman spectrometer. Table 1 summarizes all the Raman modes detected in our ferrocene crystal, along with their mode descriptions.

Figure 5 shows the FTIR spectra of our metallocene crystals measured at room temperature. In all FTIR figures, the corresponding Raman spectra are provided again to compare the correlation between Raman and IR mode numbers and energies. As is the case with the Raman measurement, the IR modes of a ferrocene crystal are clearer than those of the other two types of crystal. First, we mainly observed the intra-ring modes, which have been confirmed to be active in the molecular phase in the previous studies [21,23,29,31]. The individual inter-ring modes of $\nu_6$, $\nu_{11}$, $\nu_{21}$, and $\nu_{22}$ were hidden behind our instrumental limits, however $\nu_{11}$, $\nu_{21}$, and $\nu_{22}$ have participated in our spectra as overtone or combination modes such as $3\nu_{21}$, $\nu_4 + \nu_{11}$, $\nu_{11} + \nu_{13}$, $\nu_8 - \nu_{22}$, and $\nu_{22} + \nu_{30}$. Other interesting modes are IR modes that are only present in the crystal spectra, not in solution or vapor samples below the 1600 cm$^{-1}$ region. These are $\nu_5$, $\nu_{32}$, $\nu_{33}$, $\nu_{34}$, $\nu_{22} + \nu_{30}$, $3\nu_{21}$, and $\nu_{11} + \nu_{13}$, as shown in the figures. $\nu_{16} + \nu_{19}$, located at 1230 cm$^{-1}$, is also known to be present in crystalline form and it appeared as a small hump in our FTIR spectra of ferrocene and nickelocene crystals, but is not indicated in our figures. These modes are consistent with observations made in crystals grown using the solution method, as detailed in the previous reference [21]. We also observed the overtone band of metallocene in our FTIR spectra in the 1600 ~ 1750 cm$^{-1}$ region. These modes are not fundamental bands of metallocene, and they are also found in the vapor form. These modes are explained as a set of overtones involving C-H bending motions, which is visible due to the strong anharmonicity mixing in the vibrational states [23,32].

A unique set of vibrational modes within the range of 1900 ~ 3100 cm$^{-1}$, except for the modes of $\nu_8$, $\nu_{17}$, $\nu_{29}$, and $\nu_8 - \nu_{22}$ labelled in this region, also correspond to the crystalline features rather than the molecular forms. The study by W. K. Winter et al. [21] also identified 18 modes of ferrocene in this region that are highly consistent with our observations. In addition to these modes, we found extra new vibrations in our crystals, which are indicated by the red arrows. We have labeled all these peaks as '$\nu_{CR}$' inclusively in figure 5a, 5b and table 1.

It is also worth noting the correlation between the Raman and IR modes with respect to their associated symmetry and origin of the vibrational motion. For example, the $\nu_3$ Raman mode of $A_{1g}$ symmetry located at 1096.9 cm$^{-1}$, corresponds to the $\nu_{10}$ of $A_{2u}$, which is located at 1105 cm$^{-1}$ in the FTIR spectrum of a ferrocene crystal. They are associated with respective symmetric and asymmetric ring-breathing modes of a cyclopentadienyl ring with one (Raman) allowing dipoles and the other (IR) not. Their vibration energies are expected to differ slightly even in an individual molecule, and intermolecular interactions will introduce more variability in the crystalline form. We also believe that some of the unlabeled features in our IR spectra are related to IR modes that become dipole-allowed and undegenerated due to intermolecular interactions coupled with the vibrational motions of individual molecules. The correlation of Raman and IR modes observed in our ferrocene crystal is compared in table 1.

Another physical characteristic of our three crystals is that they tend to evaporate the yellow powder over time under ambient conditions, despite ferrocene crystals being known to be highly stable. The stability periods of the three crystals differ. Cobaltocene decomposes the fastest (within a day), followed by nickelocene (within a few days), and then ferrocene (over a month). This is because a metallocene with a higher number of electrons in their valence shells are typically less stable and more reactive [9,27]. Nickelocene, for example, has 20 electrons in its valence shell, compared to 18 in ferrocene. It is also expected that the greater the distance between the cyclopentadienyl ring and the metal ion, the more aromatic and reactive the rings will be, rendering them less stable in an external environment. In particular, the cobaltocene molecule, which is the most sensitive to air, has an $[Ar]3d^7$ electron configuration with an unpaired electron in the $E_{1g}$ orbital, and corresponding metal-ligand interaction is known to be weaker than the other two metallocene molecules [28]. The degradation timescale of crystals grown in a solution process may differ from that of our PVT-grown crystals.

## 3. Materials and Methods

**PVT method**: The raw metallocene materials for crystal growth were purchased from Sigma-Aldrich. They are sublimated via a local heating in a quartz tube inside a furnace with continuous flowing of ultra-high purity argon throughout the tube to carry the molecules to the lower temperature zone, where crystallization takes place. The sublimation temperature at which the raw compound vaporized was approximately 160 °C for all these crystals.

**LIBS**: A picosecond laser system (EKSPLA PL2231-50-SH/TH model), featuring an Nd:YAG/YVO4 laser with a 28 ps pulse duration, was employed. The sample was ablated using a 1064 nm wavelength with an energy of approximately 20 mJ, achieved by focusing the laser pulse onto the sample surface. The resulting emission light was collected via optical fiber and analyzed using a spectrometer array (StellarNet, Inc.).

**XRD**: X-Ray Diffraction measurements were conducted using a Rigaku Smartlab Diffractometer. Cu-$\alpha$ rays of wavelength 1.54 Å were employed to conduct Bragg-Brentano and Parallel beam scans in the range 2θ from 5 to 120 degrees. Scans were then analyzed using PDXL2, Vesta (cif file 2101993 on crystallography.net).

**Raman spectroscopy**: Raman spectra were collected using a Horiba iHR550 imaging spectrometer coupled to an Olympus BX41 microscope. For Raman excitation, a Near-infrared (NIR) 785 nm diode laser (iBeam-smart-785-S-WS, TOPTICA Photonics, Munich, Germany) was used. Spectra were acquired with an output power of 10 mW, using a 10× objective lens and a 600 grooves/mm grating blazed at 500 nm.

**FTIR**: Fourier-transform infrared spectra were recorded in the range of 400 ~ 4000 cm$^{-1}$ using a Thermo Nicolet Nexus 670 spectrometer (Thermo Fisher Scientific, Waltham, MA, USA). The instrument was equipped with a Nicolet iTX ZnSe attenuated total reflectance (ATR) accessory, and all measurements were performed in ATR mode at a spectral resolution of 4 cm$^{-1}$. To minimize interference from atmospheric water vapor and $CO_2$, the spectrometer was purged with dry air prior to both background and sample measurements.

## 4. Conclusions

We characterized the physical and vibrational properties of metallocene crystals grown using the PVT method. As-grown samples clearly showed the presence of the corresponding metal elements in each of our metallocene crystals (ferrocene, nickelocene and cobaltocene), as confirmed by LIBS measurements. XRD studies verified that these metallocene single crystals all have a monoclinic lattice structure with uniform morphology at room temperature. High-quality crystal samples with fewer impurities and defects, grown using this PVT method, provided clean and intense Raman and IR responses. These responses evidenced the inter- and intra-ring vibrational modes of the metallocene crystals, as well as additional vibration modes associated with their crystalline form. This study enhances our understanding of the intrinsic properties of metallocene crystals, which is important given the growing interest in using these crystalline materials in advanced technologies such as organic displays and molecular quantum information materials.

**Author Contributions**: BL conceived the idea and designed the experiments. All authors participated in preparing the sample and conducting the experiments. BL wrote the paper with contributions from other authors. All authors have accepted responsibility for the entire content of this manuscript and consented to its submission to the journal, reviewed all the results and approved the final version of the manuscript.

**Funding**: For funding support, we thank the National Science Foundation Award No. ECCS- 2347586. This work was also supported by SIUC's new faculty start-up fund and an Energy Boost Seed grant from the SIU Advanced Energy Institute.

**Data Availability Statement:** The data that support the findings of this study are available from the corresponding author upon reasonable request.

**Declaration of competing interest**: The authors declare that they have no known competing financial interests or personal relationships that may have influenced this work.

**Figures and Tables**

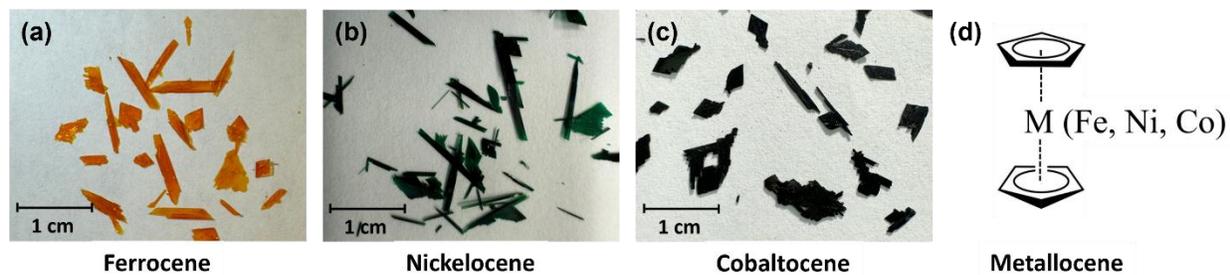

**Figure 1.** Photo images and molecular structure of metallocene crystals grown by the PVT method. (a) ferrocene, (b) nickelocene, (c) cobaltocene. (d) Molecular structures of metallocene molecules

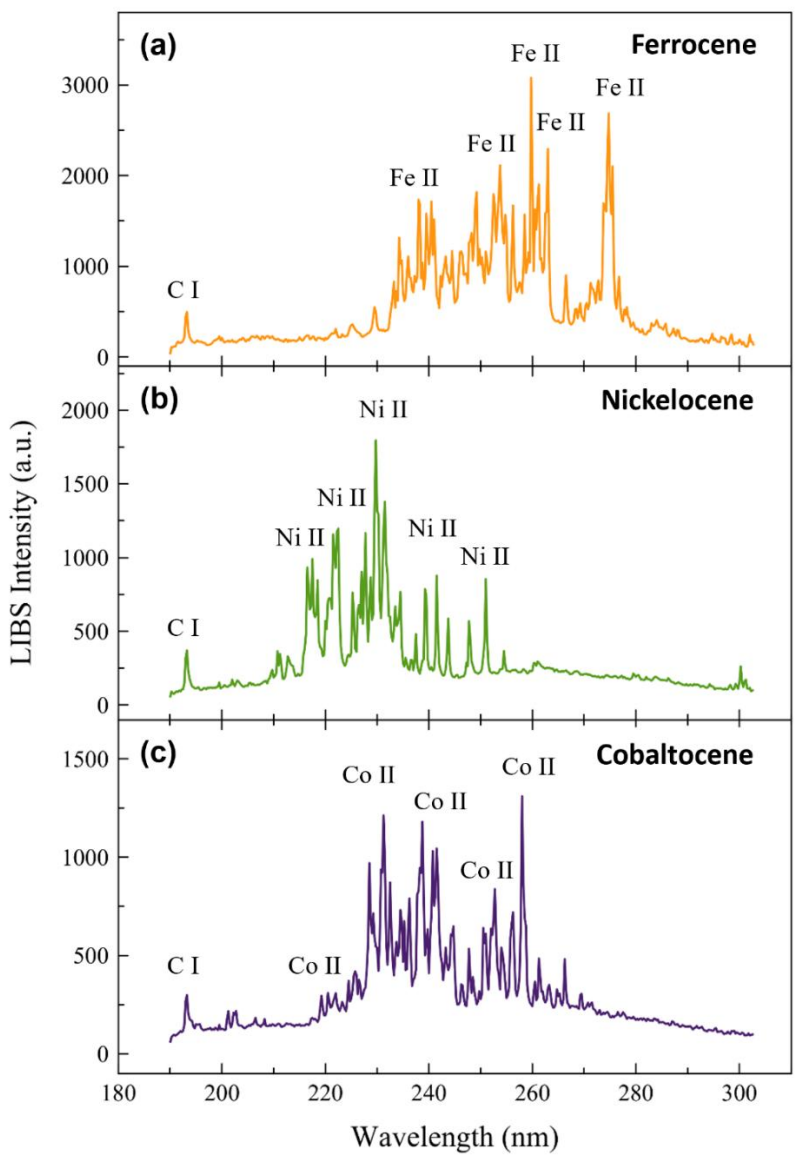

**Figure 2:** LIBS spectra for metallocene crystals. (a) ferrocene, (b) nickelocene, (c) cobaltocene.
The LIBS spectrum presented in these figures verified characteristic atomic emission lines of (a) Fe, (b) Ni, and (c) Co, respectively.

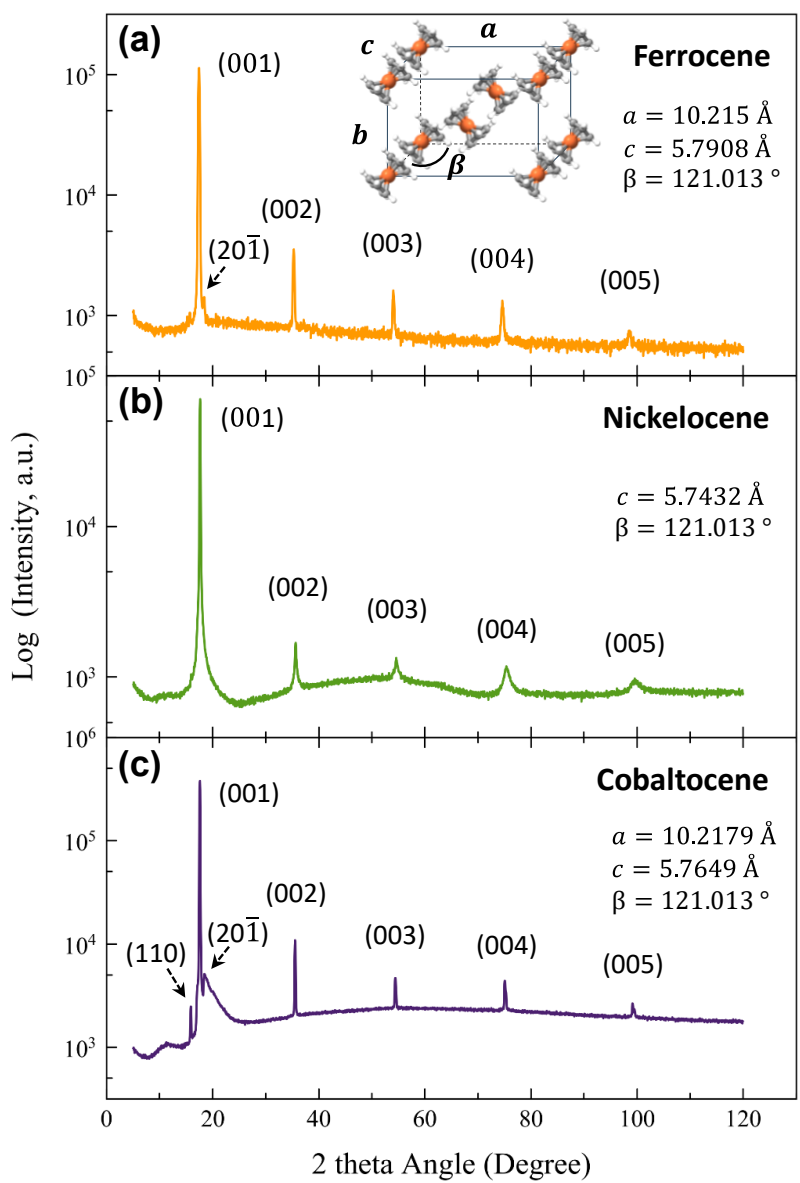

**Figure 3:** XRD data for the metallocene single crystals (a) ferrocene, (b) nickelocene, (c) cobaltocene. All samples were selected from the thin and wide single crystals with high uniformity and homogeneity to maintain their consistency and quality of the crystals.

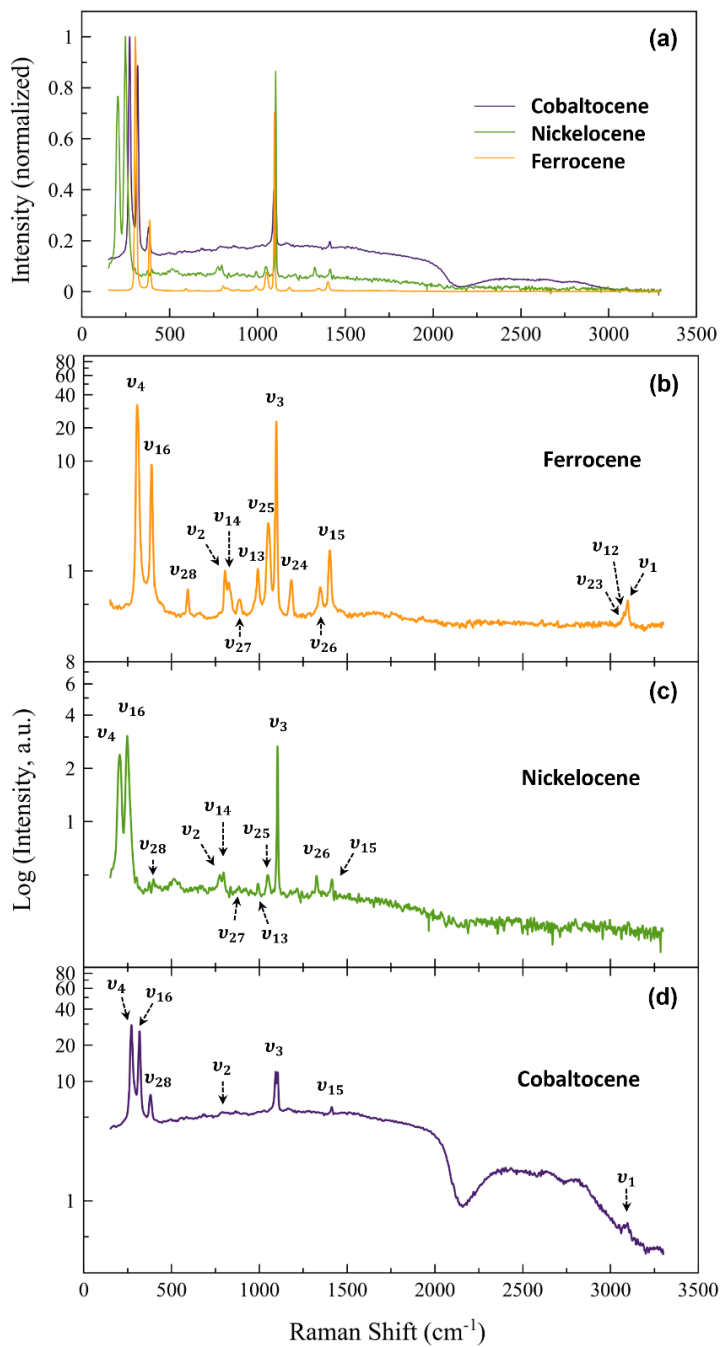

**Figure 4:** Raman spectra of the metallocene single crystals measured at room temperature. Raman spectra for (a) all metallocene, (b) ferrocene, (c) nickelocene, (d) cobaltocene single crystals. The wavelength and the powers of the excitation laser are 785 nm and $P_{exc}$ = 20 mW for all samples.

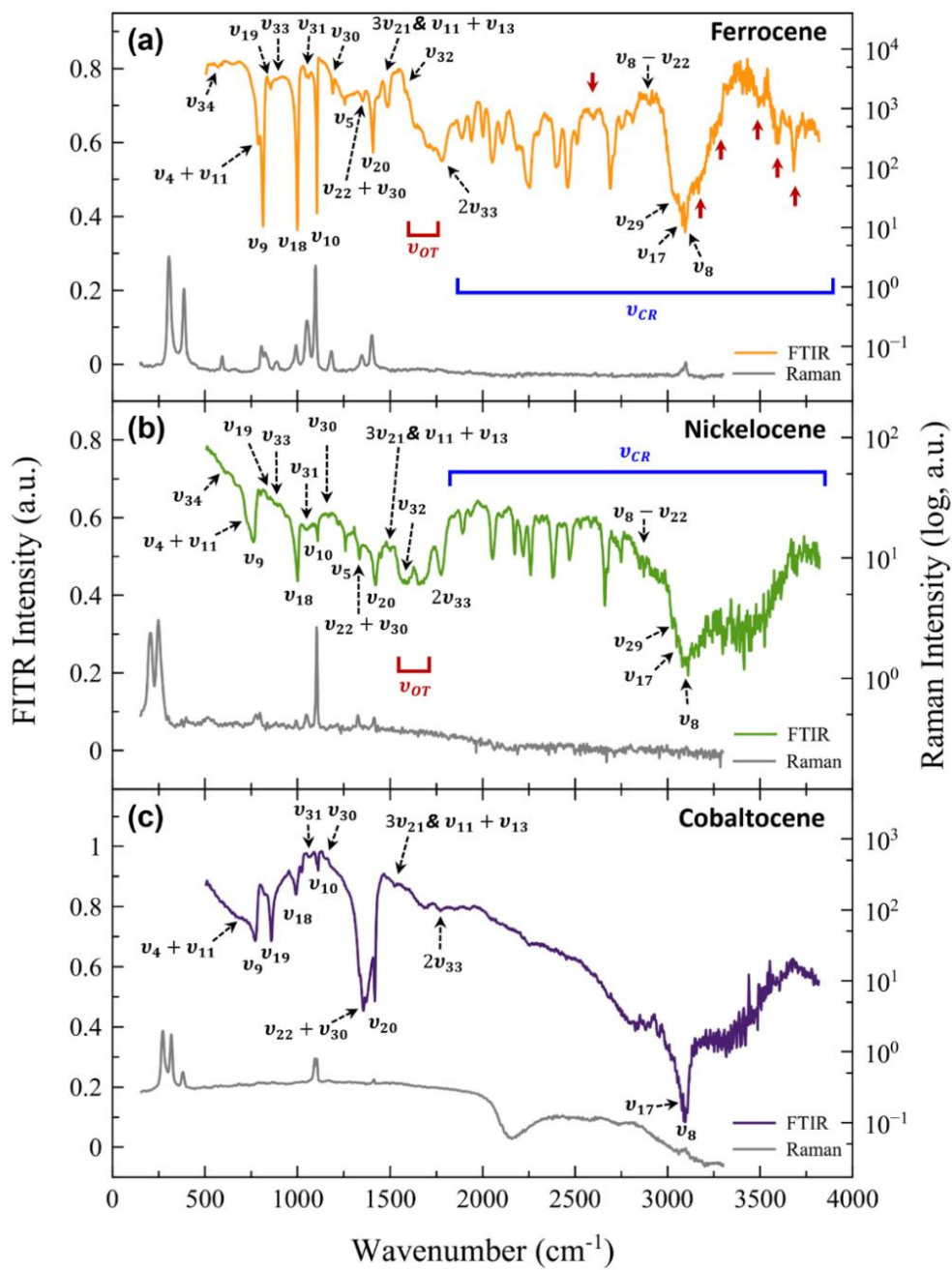

**Figure 5:** FTIR spectra of the metallocene single crystals measured at room temperature. (a) ferrocene, (b) nickelocene, (c) cobaltocene single crystals. Each figure includes the corresponding Raman spectrum (gray color) to compare between Raman and IR modes.

| Frequency assignment | Raman (cm⁻¹) | Symmetry | Description | Frequency assignment | FTIR (cm⁻¹) | Symmetry | Description | Frequency assignment | FTIR (cm⁻¹) | Symmetry |
|---|---|---|---|---|---|---|---|---|---|---|
| $\nu_1$ | 3089.5 | | Sym. C-H stretch | $\nu_8$ | 3101 | | C-H stretch | $\nu_4 + \nu_{11}$ | 788.8 | $A_{2u}$ |
| $\nu_2$ | 804.1 | | Sym. C-H bend (⊥) | $\nu_9$ | 813.8 | | C-H bend (⊥) | $\nu_{22} + \nu_{30}$ | 1351.9 | $E_{1g}$ |
| $\nu_3$ | 1096.9 | $A_{1g}$ | Sym. ring breath | $\nu_{10}$ | 1105 | $A_{2u}$ | Antisym. ring breath | $3\nu_{21}$ & $\nu_{11}+\nu_{13}$ | 1488.8 | $E_{1u}$ |
| $\nu_4$ | 304.8 | | Sym. M-ring stretch | $\nu_{11}$ | Out of range | | Antisym. M-ring stretch | $2\nu_{33}$ | 1781.9 | $E_{1g}$ |
| $\nu_7$ | missing | $A_{2g}$ | C-H bend (‖) | $\nu_5$ | 1255.5 | $A_{1u}$ | C-H bend (‖) | $\nu_{OT}$ | 1600 - 1750 | |
| | | | | $\nu_6$ | Out of range | | ring-M-ring torsion | $\nu_{15} + \nu_{21}$ | 1888 | $A_{2u}$ |
| $\nu_{12}$ | 3077.3 | | C-H stretch | $\nu_{17}$ | 3079.8 | | C-H stretch | $\nu_{14} + \nu_{10}$ | 1940.1 | $E_{1u}$ |
| $\nu_{13}$ | 991.7 | | C-H bend (‖) | $\nu_{18}$ | 999 | | C-H bend (‖) | $\nu_{13} + \nu_{18}$ | 2001.8 | $A_{2u}$ |
| $\nu_{14}$ | 826.3 | $E_{1g}$ | C-H bend (⊥) | $\nu_{19}$ | 854.3 | $E_{1u}$ | C-H bend (⊥) | $\nu_{25} + \nu_{18}$ | 2051.9 | $E_{1u}$ |
| $\nu_{15}$ | 1401.4 | | Sym. C-C stretch | $\nu_{20}$ | 1407.8 | | Antisym. CC stretch | $\nu_{13} + \nu_{10}$ | 2107.8 | $E_{1u}$ |
| $\nu_{16}$ | 386.1 | | Sym. ring tilt | $\nu_{21}$ | Out of range | | Antisym. ring tilt | $\nu_3 + \nu_{10}$ | 2185 | $A_{2u}$ |
| | | | | $\nu_{22}$ | Out of range | | ring-M-ring bend | $\nu_{15} + \nu_{19}$ | 2244.8 | $A_{2u}$ |
| $\nu_{23}$ | 3097.8 | | C-H stretch | $\nu_{29}$ | 3049 | | C-H stretch | $\nu_{29} - \nu_{14}$ | 2302.6 | $E_{1u}$ |
| $\nu_{24}$ | 1183 | | C-H bend (‖) | $\nu_{30}$ | 1191.8 | | C-H bend (‖) | $\nu_{15} + \nu_{18}$ | 2402.9 | $A_{2u}$ |
| $\nu_{25}$ | 1050.8 | $E_{2g}$ | C-H bend (⊥) | $\nu_{31}$ | 1047.2, 1060.7 | $E_{2u}$ | C-H bend (⊥) | $\nu_{15} + \nu_{31}$ | 2460.8 | $E_{1u}$ |
| $\nu_{26}$ | 1347.4 | | C-C stretch | $\nu_{32}$ | 1583.3 | | C-C stretch | $\nu_3 + \nu_{20}$ | 2512.8 | $E_{1u}$ |
| $\nu_{27}$ | 887.3 | | ring distortion (‖) | $\nu_{33}$ | 885.2 | | ring deformation (‖) | new | 2597.7 | |
| $\nu_{28}$ | 592.8 | | ring distortion (⊥) | $\nu_{34}$ | 572.8 | | ring deformation (⊥) | $\nu_{17} - \nu_{16}$ | 2692.2 | $A_{2u}$ |
| | | | | | | | | $\nu_{17} - \nu_4$ | 2757.8 | $E_{1u}$ |
| | | | | | | | | $\nu_{15} + \nu_{31} + \nu_{16}$ | 2811.7 | $E_{1u}$ |
| | | | | | | | | $\nu_8 - \nu_{22}$ | 2923.6 | $E_{1g}$ |
| | | | | | | | | new | 3166.6 | |
| | | | | | | | | new | 3266.9 | |
| | | | | | | | | new | 3506 | |
| | | | | | | | | new | 3590.8 | |
| | | | | | | | | new | 3681.5 | |

The $\nu_{CR}$ label applies to the block of combination/difference modes in the rightmost section.

**Table 1:** Description of the Raman and IR vibration modes measured in the ferrocene single crystals grown by the PVT method.